# Pressure-induced mixed states caused by spin-elastic interactions during first-order spin phase transition in spin crossover compounds

Ruixin Li, Viktor M. Kalita,* Hennagii Fylymonov, Wei Xu, Quanjun Li,* Jose Antonio Real, Bingbing Liu, Georgiy Levchenko*

**ABSTRACT:** Recently, the possibility of exploiting the phenomenon of spin transition (ST) has been intensively investigated, therefore, it is particularly important to study the behavior of ST under various stimuli. Here, the shape and content of the intermediate phase of ST in Hoffmann-like compounds $[Fe(Fpz)_2M(CN)_4]$(M = Pt, Pd) under external stimuli are studied. For this purpose, magnetic and Raman spectroscopy measurements were carried out. In pressure-induced spin transition (PIST), a mixture of high-spin and low-spin states appears, while in temperature-induced spin transition (TIST), a homogeneous state occurs. The first-order ST induced by pressure has a hysteresis, but is not abrupt. Whereas, the temperature-induced spin transition at ambient pressure is hysteretic and abrupt. To investigate this difference, we discuss using a thermodynamic model that considers elastic interactions, showing that the slope of the hysteresis loop is related to the appearance of internal pressure, which is related to the difference in sample compressibility under high spin and low spin states.

# INTRODUCTION

Spin crossover (SCO) complexes are the objects of intense research for almost five decades, thanks to their attractive switching ability.[1] They can switch between a low spin (LS) and a high spin (HS) states if external stimuli such as: temperature, pressure, light, X-Ray or electric and magnetic fields are applied.[2-3] The switching leads to modifications in physical properties of SCO materials conferring a high potential in applications as saving information materials, sensors or new generation of devices. Among all types of spin transition (ST), the first-order spin crossover, which has a large hysteresis width and a transition temperature $T_{1/2}$ at room temperature, has the most potential for application. Since most ST cannot directly meet the actual application, it is an urgent task to study the possibility of precise regulation of ST.[4-15] The use of both various external physical stimuli, such as temperature, pressure,[6-8, 16-21] light,[22-29] magnetic field,[30-31] X-ray radiation,[32] and chemical, such as the structural modification, replacement of coordinated metal ligands, organic radicals or metals,[33-35] which can change the properties of the complex in a wide range, and then regulate the ST phenomenon.

Among the external stimuli, the pressure plays an exceptional role in studying the physics of the ST phenomenon, since it can directly change the crystalline field and interionic interactions. In addition, the pressure allows the SCO compound to obtain the necessary characteristics required for practical applications, that is, the transition temperature $T_{1/2}$ is at room temperature and a large hysteresis width.

Despite the widespread study of the effect of temperature and pressure on spin transition, most attention was directed for studying the change in the transition temperature and the width of the hysteresis under pressure. No attention has been paid to the shape of the hysteresis loop of the first-order ST, which can be rectangular or inclined in the shape of parallelogram-like. However, the loop shape is important for understanding the thermodynamics of the phase transition. In the case of ST with a rectangular loop, one-step transition in the form of a jump occurs. If the loop has a parallelogram-like shape, it must be assumed that there is gradual transition between the phases in this case, the problems with the phase and spin state observed in ST systems with tilted hysteresis arise. The shape of the hysteresis loop of the ST and the possibility of mixed states formation are very important when studying the conditions for the use of spin phase transition, as well as when studying the physics of this phenomena under pressure. In this article, we will show that the formation of the mixed states is not associated with the non-equilibrium phase transition, and it is a consequence of the thermodynamic stability of the crystal subjected to compression.

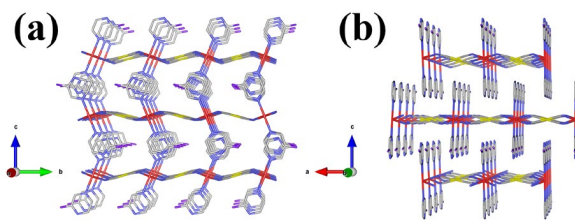

Figure. 1. The structure of [Fe(Fpz)$_2$M(CN)$_4$], where M = Pt, Pd, shows the stacking of three consecutive layers.

For this, the recently synthesized compounds of Hoffmann type [Fe(Fpz)$_2$M(CN)$_4$], where M is Pt and Pd, are a very convenient objects.[36] In the presented work, the effect of the pressure on the ST in these compounds was investigated by measuring the temperature dependences of the magnetic moment under pressure as well as the Raman spectroscopy measurements under variation of the temperature and pressure. The equipment and experimental methods used in this work are described in the supporting information.

## RESULTS

The samples [Fe(Fpz)$_2$Pd(CN)$_4$] and [Fe(Fpz)$_2$Pt(CN)$_4$] have been synthesized according to ref 36. These compounds are isostructural and show the orthorhombic Pmna space group. The Fe(II) atoms lie at an inversion centers which define an elongated octahedral coordination site FeN$_6$. Axial positions are occupied by the nitrogen atoms of fluorinated pyrazine ligands. The equatorial positions are occupied by the nitrogen atoms of four equivalent centrosymmetric square-planar [M(CN)$_4$]$^{2-}$ groups, where M is Pd, Pt. The layers consisting iron and M ions stack one on top each other in such a way that the axial ligands of one-layer point to the center of the {Fe$_2$[M(CN)$_4$]$_2$} square windows of the adjacent layers, result in the Fpz ligands of consecutive layers are interdigitated.

Figure 2(a) shows the temperature dependences of the fraction of high spin state ($\gamma_{HS}$) determined from magnetic measurements at various external pressures for [Fe(Fpz)$_2$Pt(CN)$_4$] (Fe-Pt), and Figure. 2(b) shows the same data for [Fe(Fpz)$_2$Pd(CN)$_4$] (Fe-Pd). At ambient pressure the hysteresis has a rectangular shape, which meets the abrupt conversion between low spin and high spin states. Under pressure, the hysteresis loops become smoother and inclined. The occurrence of the inclination of the hysteresis loop and the appearance at the beginning the parallelogram-like, and then the S-shape loops means that under pressure abrupt transition between two homogeneous spin phases does not occur, but in the transition area a mixed state is formed. The Fe-Pt and Fe-Pd phase diagrams, which are qualitatively consistent, are shown in Figure S1.

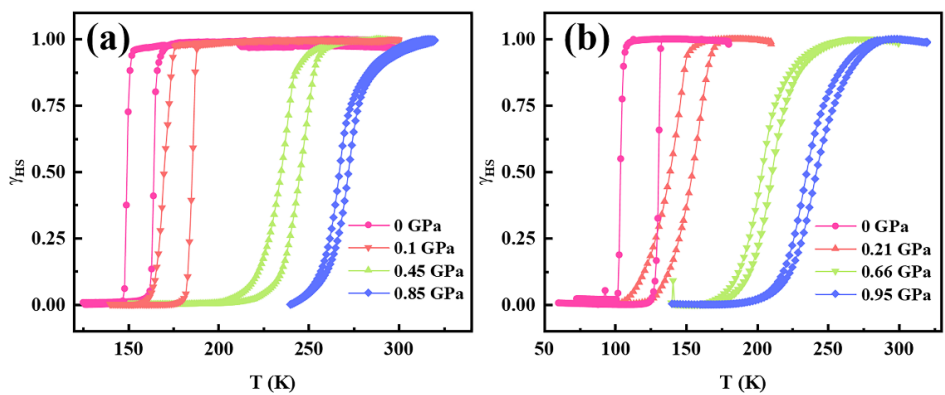

Figure. 2. The temperature dependences of the high spin fraction of the molecules $\gamma_{HS}$ at various external pressures: (a) for [Fe(Fpz)$_2$Pt(CN)$_4$], and (b) for [Fe(Fpz)$_2$Pd(CN)$_4$]

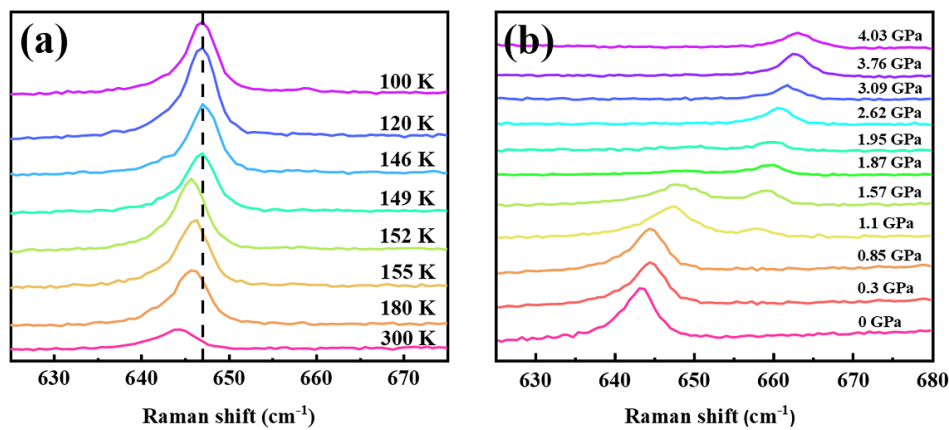

Figure. 3. The evolution of the in-plane bending vibrational modes of F-pyrazine ligands: (a) upon decreasing temperature mode and (b) upon increasing pressure mode at room temperature for Fe - Pt.

The transition temperature gradually increases with increasing pressure, but deviates slightly from linearity at higher pressures. The hysteresis widths of the two compounds decrease with increasing pressure at low pressures, but exhibit completely different behaviors at high pressures. In the Fe-Pt diagram, the hysteresis width drops to zero at P = 0.7 GPa, but then increases and begins to fall again. In the Fe-Pd phase diagram, the hysteresis decreases to the value $\Delta T_{1/2} = 7$ K at P = 0.39 GPa and then remains constant. Theoretical predictions show that the hysteresis width will always remain zero after the spin transition under pressure is turned into spin crossover. However, experiments have found that at the critical point of spin transition and spin crossover, the system is not stable, will cause different phenomena to occur. To finally elucidate the nature of this phenomenon, structural measurements with simultaneous pressure and temperature control are necessary.

Figure 3a presents the evolution of the in-plane bending vibrational mode of F-pyrazine ligands upon decreasing the temperature and Figure 3b presents the same vibration upon increasing the pressure at room temperature of the Fe-Pt received previously in ref 37. Similar to the compound Fe-Pt, the evolution of the Raman vibration of Fe-Pd with temperature and pressure is shown in Figure S2.

As has been shown in ref 37-39,[37-39] the in-plane vibration mode can be used to track pressure-induced spin transition. From the evolution of Raman spectrum with pressure, we observe that the vibration modes corresponding to the high spin states of the compounds Fe-Pt and Fe-Pd are located at 644 cm$^{-1}$ and 643.5 cm$^{-1}$, respectively. When the pressure increases, the mode related to the HS state shifts to a high wave number and the intensity decreases. At the same time, as the pressure increases, a new vibration mode corresponding to the LS state appears (658 cm$^{-1}$ for Fe-Pt and 659 cm$^{-1}$ for Fe-Pd), and it can be seen that the HS and LS modes coexist. As the pressure increases, the intensity of the newly emerging modes increases, and the intensity of the modes associated with the HS state decreases. Under sufficiently high pressure, the HS state mode disappears and only the LS state mode remains. The Raman spectra of TIST is completely different from the spectra of PIST. When the temperature decreases, the HS mode shifts to the high frequency region, but this shift is very small (Fe-Pt from 644 to 647.5 cm$^{-1}$, Fe-Pd from 643.5 to 645.5 cm$^{-1}$), and there is no two-phase coexistence. Such behavior of Raman modes reflects the significant difference in TIST and PIST. Under pressure, the hysteresis loop of TIST becomes tilted (Figure 2), which means that pressure will also cause mixed LS and HS states in TIST.

Therefore, it can be seen from the experimental data that the pressure changes the phase transition process of ST. The pressure causes a mixed state at TIST, changing the slope of the hysteresis loop, as shown in Figure 2. At PIST, the pressure dependence of the high spin state is shown in Figure S3.

## DISCUSSION

The presented data on the study of TIST in the case of a first order phase transition demonstrate sharp transitions with a rectangular hysteresis.[9, 34, 40-42] Under pressure, the transition temperature moves to high temperature, and the steepness of the transition decreases and becomes inclined. These features are presented in Figure 2. The displacement of the transition and its steepness depends on the composition and pressure. In some cases, such as in ref 41,[41] the change in the steepness is a very small.

Spin crossover is a consequence of the competition of the crystal field and effective interionic interaction. The increasing of the first leads to the continuous spin transition - the thermodynamic path is bypassing the tricritic point and goes past the critical lines, and the increasing of the second one leads to the hysteresis with an intermittent change in the order parameter - the thermodynamic path crosses the critical lines of the stability of the low spin and high spin phases.

The displacement of the transition temperature induced by pressure can be explained by the increase in the magnitude of the crystal field during compression, but the formation of two phases is impossible. The formation of the two phase is associated with an additional interionic interaction arising from crystal compression.

To explain observed difference between TIST and PIST, we analyzed these transitions using thermodynamic approximation. In Figure 4 the dependence of the order parameters of the homogeneous low spin and high spin states on the normalized temperature $t = T/Tc$ for TIST, is shown according to ref 42.[42] The low-spin part of order parameter exists for $t < t_{LS}$ and it's derivative $d\gamma_{HS}(t)/dt > 0$. A high-spin branch of the solution $\gamma_{HS}(t)$ exists for $t > t_{HS}$ and for it $d\gamma_{HS}(t)/dt > 0$ as well. The branch of the solution in the interval $t_{HS} < t < t_{LS}$ with $d\gamma(t)/dt < 0$ is unstable.

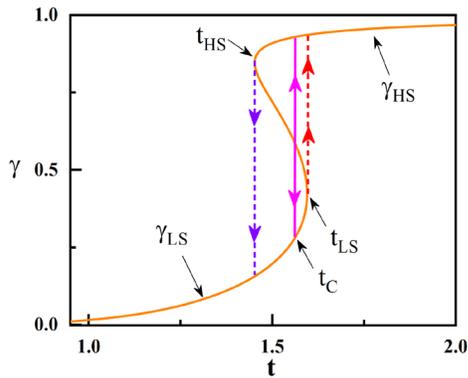

Figure. 4. The dependence of the order parameter on the normalized temperature t for TIST. $t_C$ is the temperature of the ST, $t_{LS}$ is the temperature of the LS state stability, $t_{HS}$ is the temperature of the HS state stability.

At the point of ST $t_C$, the energy of the LS state is equal to the free energy of the HS state $f_{LS}(t_C) = f_{HS}(t_C)$. The order parameter $\gamma$ is a scalar, a change in its value does not lead to a change in symmetry, therefore this is an isostructural phase transition of the first order.[42]

At nonequilibrium transition between LS and HS states, hysteresis is possible. Upon transition from the HS to the LS, it's left boundary $t_{HS}$, is indicated by vertical dashed line with down arrows. Upon transition from LS to HS, the right hysteresis boundary is indicated by vertical line with up arrows. Hysteresis is limited by two points $t_{HS}$ and $t_{LS}$. At these temperatures the reduced order parameter dumps from one to zero at $t_{HS}$ and from zero to one at $t_{LS}$, respectively. So, with temperature decrease the HS is stable down to $t_{HS}$, with temperature increase the LS state is stable up to $t_{LS}$ and the two phases are never mixed inside a hysteresis loop.

Under pressure, first-order spin transitions with hysteresis, but not always sharp transitions.[9, 37, 40] To our knowledge, the gradual transitions with hysteresis were not described and reason of the appearance and increasing inclination with increasing pressure also is not clarified. The models describing the ST consider the pressure influence on the order parameter, transition temperature and hysteresis width,[37, 40-41, 43-46] but do not describe the pressure influence on the slop and content of the ST hysteresis loop. For doing it, we consider the contribution of pressure on the free energy of the system.

**Theoretical consideration**

*Free energy, equation of state and spontaneous deformation.* When describing the first order ST, the expression for Gibs free energy can be represented in a traditional form $G = E - TS$ [43], where E is enthalpy $\left(E = \Delta\gamma - \frac{1}{2}J\gamma^2\right)$ and S is entropy ($S = -\gamma \ln \gamma - (1 - \gamma) \ln(1 - \gamma) + \gamma \ln g_{HS} + (1 - \gamma) \ln g_{LS}$). The equilibrium state is obtained from the equation $dG/d\gamma = 0$ and has the form:

$$\Delta - J\gamma - T \ln g - T \ln \frac{1-\gamma}{\gamma} = 0 \qquad (1)$$

where Δ is a splitting energy between the high spin and low spin states, characterizes the effect of the crystal field, J is a parameter characterizing molecular interaction, γ is the order parameter, $g = g_{HS}/g_{LS}$ and $g_{HS}$, $g_{LS}$ are the degenerations of the high spin and low spin states.

The equation of state (Eq. (1)) well describes the rectangular type $\gamma - T$ hysteresis loop of the ST when heating or cooling the sample, and it is often used for this purpose in the literature.[42, 44-46]

Since Eq. (1) does not contain elastic terms, it cannot describe PIST and γ-P hysteresis. For that the contribution of the pressure to the free energy should be considered. This contribution depends from volume deformation of the sample $\delta = \Delta V/V$ caused by pressure which equals to relative change of cell volume with one magnetic ion. In linear approximation one can write $\Delta(\delta) = \Delta_0 + \alpha\delta$, $J(\delta) = J_0 + \beta\delta$, where $\Delta_0 = \Delta(\delta = 0)$, $\alpha = (\partial\Delta(\delta)/\partial\delta)_{\delta=0}$, $J_0 = J(\delta = 0)$, $\beta = (\partial J(\delta)/\partial\delta)_{\delta=0}$.[42]

Compression accounting leads to a change in expression for inner energy

$$E(\gamma, \delta) = \Delta_0\gamma - \frac{1}{2}J_0\gamma^2 + \alpha\delta\gamma - \frac{1}{2}\beta\delta\gamma^2 + \frac{1}{2}K\delta^2 \quad (2)$$

The third and fourth components in Eq. (2) characterizes the spin-elastic contribution, which is considered in ref 43. From the condition $p = -\partial E/\partial\delta$ we get

$$\delta = \frac{1}{K}(-p_e - \alpha\gamma + \frac{1}{2}\beta\gamma^2) \quad (3)$$

where $K$ is bulk modulus. External positive pressure $p_e > 0$ leads to compression with $\delta < 0$.

It can be seen from Eq. (3) that the compression leads to a linear according to order parameter addition to the pressure, and the interionic interaction gives a quadratic from the value of the order parameter to the pressure. From Eq. (3) it follows that the spin-elastic addition to pressure leads to a violation of the Hooke's law.

When considering only the first order of $\gamma$ smallness and the assumption that $|\alpha| >> |\beta|$, the $\delta = \frac{1}{K}(-p_e - \alpha\gamma)$ and at $p_e = 0$ the change in volume caused only by the spin-elastic interaction in a high spin state $\gamma = \gamma_{HS}$ and low spin state $\gamma = \gamma_{LS}$ will be equal to

$$\delta_{HS} = -\frac{1}{K}\alpha\gamma_{HS}, \quad \delta_{LS} = -\frac{1}{K}\alpha\gamma_{LS} \quad (4)$$

where $\delta_{HS}$ or $\delta_{LS}$ are the spontaneous deformations of high and low spin states, respectively. Their difference equals to change of the relative value of the cell volume at the spin phase transition. At $p_e = 0$ in the high spin state $\delta_{HS} > 0$ and $\delta_{HS} > \delta_{LS}$, therefore $\alpha < 0$.

Substituting $\delta = \frac{1}{K}(-p_e - \alpha\gamma)$ in Eq. (2), the expression of pressure influence on inner energy is received in the form:

$$E(\gamma, p_e) = \Delta_0\gamma - \frac{1}{2}(J_0 - \frac{\alpha^2}{K})\gamma^2 + \frac{p_e^2}{2K} \quad (5)$$

The pressure increases the crystalline field and consequently the splitting of $t_{2g}$ and $e_g$ levels inducing the low spin state at critical pressure $p_{cr}$.

In the case of existence of the nonequilibrium states, two critical pressures $p_{cr}^{(+)}$ and $p_{cr}^{(-)}$ exist and the hysteresis loop should be rectangular.

So, the consideration of the elastic interactions explains the nature of induced ST, however cannot explain the inclined form of the $\gamma - P$ transition.

***Mixed state.*** As shown in Figure S3, the experimental pressure induced spin transitions for both compounds have the hysteresis, therefore they are the first order transitions. However, the transition curves are inclined. This is an unusual for simultaneous (abrupt) transition of the all sample. The inclining of transition curves means that the induced ST occurs through a mixed state in which one part of the ions continues in a high spin state, and the other part passed into the low spin state. If denote by $c$ the concentration of ions in high spin state then the concentration of the low spin state ions equals to (1-$c$). In a mixed state, the average value of the order parameter will be equal $\bar{\gamma} = c\gamma_{HS} + (1-c)\gamma_{LS}$ and the average value of the volume caused by the spin-elastic coupling in the mixed state will be equal to

$$\bar{\delta} = -\frac{1}{K}\alpha(c\gamma_{HS} + (1-c)\gamma_{LS}) \text{ or } \bar{\delta} = -\frac{\alpha}{K}\bar{\gamma} \quad (6)$$

To find the bond between pressure and the order parameter, we considered the second order of smallness of the spin-elastic contribution to the crystal energy. Using the approach in ref 44 the deposit into energy recorded in the second order of smallness by volume can be represented as:

$$E_{HS-LS} = D(\frac{\bar{\delta}}{\delta_{LS}} - 1)(1 - \frac{\bar{\delta}}{\delta_{HS}}) \quad (7)$$

where $\delta_{LS} < \bar{\delta} < \delta_{HS}$ and D - constant.

Indeed, using Eq. (6), the expression for energy (Eq. 7) can be submitted by the form of a product of concentrations of both phases, $E_{HS-LS} \sim c(1-c)$, which is alleged in ref 44 is the term of the second order of smallness in comparison with the enthalpy of the sample.

From $-\partial E_{HS-LS}/\partial\bar{\delta}$ we find the additional pressure in the sample in a mixed state.

$$\Delta p = -\partial E_{HS-LS}/\partial\bar{\delta} = -D(\frac{1}{\delta_{LS}} + \frac{1}{\delta_{HS}} - 2\frac{\bar{\delta}}{\delta_{LS}\delta_{HS}}) \quad (8)$$

At the limit points of the mixed state $\bar{\gamma} = \gamma_{HS}$, $\bar{\gamma} = \gamma_{LS}$ in which $\bar{\delta}(\bar{\gamma} = \gamma_{HS}) = \delta_{HS}$, $\bar{\delta}(\bar{\gamma} = \gamma_{LS}) = \delta_{LS}$ the additional pressures equal to

$$\Delta p_1 = \Delta p(\bar{\gamma} = \gamma_{HS}) = \Delta p(\bar{\delta} = \delta_{HS}) = D\frac{\delta_{HS} - \delta_{LS}}{\delta_{LS}\delta_{HS}} \quad (9)$$

$$\Delta p_2 = \Delta p(\bar{\gamma} = \gamma_{LS}) = \Delta p(\bar{\delta} = \delta_{LS}) = -D\frac{\delta_{HS} - \delta_{LS}}{\delta_{LS}\delta_{HS}}$$
$$(10)$$

At $D < 0$ the values $\Delta p_1 < 0$, $\Delta p_2 > 0$. If modules $|\Delta p_1| = |\Delta p_2| = \Delta p_\kappa$, then $\Delta p_1 = -\Delta p_\kappa$, and $\Delta p_2 = \Delta p_\kappa$. At D <0 the energy (Eq. 7) is negative and this sign corresponds to the sign adopted in the ref 44 model describing mixed states with high and low spin.

With Eq. (8), it is possible to obtain that in a mixed state, the spin-elastic contribution to the inner pressure linearly depends on the value of the average order parameter:

$$\Delta p(\bar{\gamma}) = \frac{-\Delta p_\kappa}{\gamma_{HS} - \gamma_{LS}}(2\bar{\gamma} - \gamma_{LS} - \gamma_{HS}) \quad (11)$$

Additional pressure $\Delta p(\bar{\gamma})$ will change the view of the hysteresis loop, making it inclined.

At the same time, we obtain that the changing of the crystal field value by pressure is able to mix spin phases, without changing the type of the hysteresis loop. To obtain a parallelogram-like loop, it is necessary to consider an additional interionic interaction that phenomenologically explains the relationship between the value of HS fraction and pressure.

According of this consideration in the case of a mixed state during the transition from a high spin to the low spin state the critical pressure equals to the external pressure, from which the additional internal pressure is subtracted, $p_{cr}^{(+)} = p_e - \Delta p$ or $p_e = p_{cr}^{(+)} + \Delta p$, where $p_{cr}^{(+)}$ is the critical pressure at HS to LS transition, $p_e$ is external pressure and $\Delta p$ is an additional internal pressure according to Eq. (8-10). Thus, for this branch of hysteresis we have

$$p_e(\bar{\gamma}) = p_{cr}^{(+)} - \frac{\Delta p_\kappa}{\gamma_{HS} - \gamma_{LS}}(2\bar{\gamma} - \gamma_{LS} - \gamma_{HS}) \quad (12)$$

For the other branch of the hysteresis in a mixed state when moving from a low spin to a high spin state we have $p_e(\bar{\gamma}) = p_{cr}^{(-)} - \frac{\Delta p_\kappa}{\gamma_{HS} - \gamma_{LS}}(2\bar{\gamma} - \gamma_{LS} - \gamma_{HS})$ (13)

From Eq. (12-13), it is seen that in a mixed state, the external pressure linearly depends on the order parameter with the same and constant tilt.

So, when applying the pressure at the point $\bar{\gamma} = \gamma_{HS}$, the external pressure of the transition from a high spin state into the mixed state will be less of the critical pressure and equal to $p_{cr}^{(+)} - \Delta p_\kappa$. With a further increase in pressure, the mixed state goes into a homogeneous low spin state at the point $p_{cr}^{(+)} + \Delta p_\kappa$. If the pressure decreases, the low spin state will be observed when $p_e > p_{cr}^{(-)} + \Delta p_\kappa$. When the pressure lies in the interval $p_{cr}^{(-)} - \Delta p_\kappa < p_e < p_{cr}^{(-)} + \Delta p_\kappa$, then there will be a mixed state. And finally, if the pressure is less of $p_e < p_{cr}^{(-)} - \Delta p_\kappa$, then a high spin state will be restored. The hysteresis loop, considering the additional spin - elastic interactions in Eq. (9-10), is shown in the Figure 5. It was accepted for simplicity that $\gamma_{HS} = 1$ and $\gamma_{LS} = 0$.

Based on the described concept of mixed state, all four points (vertices) of parallelogram of the loop are critical. The point $p_{cr}^{(+)} - \Delta p_\kappa$ corresponds to the continuous transition from the high spin state into a mixed state with an increase in pressure, the point $p_{cr}^{(+)} + \Delta p_\kappa$ corresponds to the continuous transition from a mixed state into a low spin state with an increase in pressure, the point $p_e = p_{cr}^{(-)} + \Delta p_\kappa$ corresponds to the continuous transition from the low spin state into a mixed state with a decrease in pressure and the point $p_{cr}^{(-)} - \Delta p_\kappa$ corresponds to a continuous transition from a mixed state to low spin state with a decrease in pressure. This cycle in the Figure 5 is indicated by arrows. Thus, in contrast to the TIST, in the PIST the pressure leads to the inclination of the hysteresis loop and to the appearance of a mixed HS and LS states. According to Eq. (S9-S10) the inclination depends on difference of volume change under pressure in HS and LS states. As this difference increases, the slope increases. Figure 6 shows the pressure induced spin transitions at room temperature for Fe-Pt and Fe-Pd obtained from Raman spectroscopy measurements. The area between the bold vertical lines refers to the intermediate state during the ideal first-order ST. Figure 6 shows the influence of the inner pressure $\Delta p_\kappa$ on spin transition of Fe-Pt and Fe -Pd compounds. It moves the beginning of the transition from point $p_{cr}^{(+)}$ to the $p_{cr}^{(+)} - \Delta p_k$ and the end of the transition from the $p_{cr}^{(+)}$ to the $p_{cr}^{(+)} + \Delta p_k$ during the increase of the pressure. At release the pressure, the beginning of the transition from point $p_{cr}^{(-)}$ moves to the $p_{cr}^{(-)} + \Delta p_k$ and the end of the transition moves from $p_{cr}^{(-)}$ to the $p_{cr}^{(-)} - \Delta p_k$. Therefore, the shape of the hysteresis loop changes from a rectangle to a parallelogram-like shape. The value of the $\Delta p_\kappa$ equals approximately to 0.68 GPa for the Fe-Pt and 1.0 GPa for Fe-Pd. The difference between these values is small and depends from the difference of compressibility of these compounds in high spin and low spin states. Using these values $\Delta p_k$, the relative change in the volume $\delta_{HS}$ (equal 0.0237 for Fe-Pt and 0.022 for Fe-Pd) obtained from the X-ray measurements in ref 36 and the modulus of elasticity in the high-spin state ( equal 7.14 GPa for Fe-Pt and 9.62 GPa for Fe-Pd), we calculated the coefficients $\alpha$ from expression Eq. (4) which determines the spin elastic interactions and the coefficients D from equation Eq. (9) or Eq. (10)

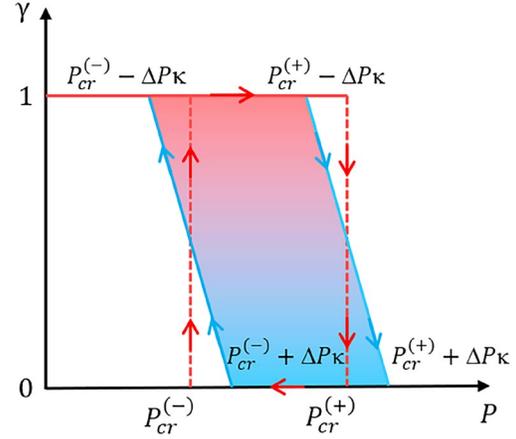

Figure. 5. The $\gamma - P$ diagram of PIST. The dotted line with arrow marks the rectangular type $\gamma - P$ hysteresis loop. The solid line marks the hysteresis loop established considering spin-elastic interactions. The slanted graph corresponds to the mixed state.

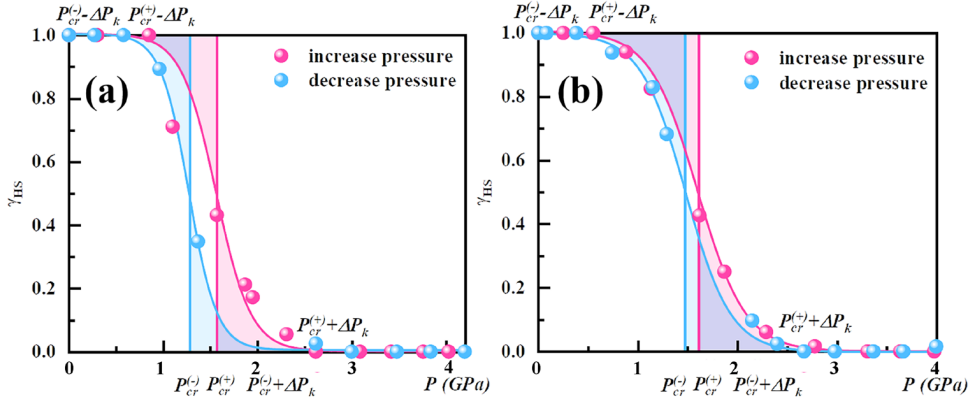

Figure. 6. The influence of the internal pressure $\Delta p_\kappa$ on pressure induced spin transition at room temperature: (a) Fe-Pt and (b) Fe-Pd.

determining the change of the spin elastic energy at ST. For Fe-Pt, α=0.17 GPa, D=0.31 GPa, and for Fe-Pd, α=0.21 GPa, D=0.5 GPa. It can be seen that as the bulk modulus increases, the spin-elastic energy increases.

In TIST at atmospheric pressure, the transitions for both compounds have rectangular form and the mixed state does not exist (Figure 2). At applying pressure, the transition becomes inclined. It means that under pressure the mixed state appears. It happens because the pressure $\Delta p$ depends on average value $\bar{\gamma}$ which changes with changing pressure and temperature. As the temperature decreases, $\Delta p$ decreases and the transition stretches to the left, and as the temperature increases, the transition stretches to the right. Therefor the transition becomes gradual with hysteresis. The pressure influence on the shape of transition loop recently has been studied in ref 47 and has been shown that the gradient of the pressure results in inclined the ST loop.[47] Here we show that not only the pressure gradient, but also the internal pressure which caused by different compressibility of HS and LS molecules are changed during the pressure variation inclines the hysteresis loop and induces the mixing intermediate state. The difference between internal and external pressure is comparable to the difference between magnetic induction inside a magnet and an external magnetic field. The dependence of the internal pressure on external pressure causes the inclination of the hysteresis width and appearance of the mixed intermediate state.

## CONCLUSION

The shape and content of the hysteresis of TIST and PIST are experimentally and theoretically investigated based on the two-dimensional Hoffmann-type compound [Fe(Fpz)$_2$M(CN)$_4$](M = Pt and Pd). In the experiment, we carried out magnetic and Raman spectroscopy measurements. In theory, we apply a thermodynamic model that considers elastic interactions. The first-order TIST is different from PIST in shape: TIST is an abrupt transition with hysteresis, while PIST is a gradual transition with hysteresis. Among them, the biggest difference between them is the composition of the intermediate state: the temperature-induced spin transition occurs without the appearance of the mixed state, and the ST caused by the pressure occurs in the mixed state. This is the first direct proof of the appearance of a mixed state in a pressure-induced spin transition. The theoretical basis shows that the essence of the mixed state is the spin-elastic interaction. Therefore, under pressure, the homogeneous state within the hysteresis loop of TIST can be transformed into a mixed state. The nature of appearance the mixed state in PIST and in TIST under pressure is explained by appearance of internal pressure during transition which creates mixed states and is caused by difference in compressibility of two LS and HS states. With increase of the bulk modulus of the compounds the spin elastic coupling increases and the value of the internal pressure increases also. The appearance of such pressure is similar to the appearance of magnetic induction (B) in magnetic materials placed in external magnetic field which does not equal to external magnetic field (H) and electric induction(D) in the sample placed in electrical field (E). Due to the influence of spin-elastic interactions on the shape and content of ST, pressure is able to switch between homogeneous and mixed states in the magnetic hysteresis, which provides a certain theoretical and experimental basis for the application of ST.

# ASSOCIATED CONTENT

# AUTHOR INFORMATION

**Corresponding Authors**

**Viktor M. Kalita** - *Institute of Magnetism of NAS of Ukraine and MES of Ukraine, 36-b Vernadsky Blvd., Kyiv 03142, Ukraine; National Technical University of Ukraine 'Igor Sikorsky Kyiv Polytechnic Institute', Prospekt Peremohy 37, Kyiv 03056, Ukraine; Institute of Physics, NAS of Ukraine, Prospekt Nauky 46, Kyiv 03028, Ukraine; Email:* vmkalita@ukr.net

**Quanjin Li** - *State Key Laboratory of Superhard Materials, International Centre of Future Science, Jilin University, Changchun 130012, China; Email:* liquanjun@jlu.edu.cn

**Georgiy Levchenko** - *State Key Laboratory of Superhard Materials, International Centre of Future Science, Jilin University, Changchun 130012, China; Donetsk Institute for Physics and Engineering named after O.O. Galkin, NAS of Ukraine, Prospekt Nauky 46, Kyiv 03028, Ukraine; Email:* g-levch@ukr.net

**Authors**

**Ruixin Li** - *State Key Laboratory of Superhard Materials, International Centre of Future Science, Jilin University, Changchun 130012, China*

**Hennagii Fylymonov** - *Donetsk Institute for Physics and Engineering named after O.O. Galkin, NAS of Ukraine, Prospekt Nauky 46, Kyiv 03028, Ukraine*

**Wei Xu** - *State Key Laboratory of Inorganic Synthesis and Preparative Chemistry, College of Chemistry, Jilin University, Changchun 130012, China*

**Jose Antonio Real** - *Institut de Ciència Molecular, Departament de Química Inorgànica, Universitat de València, E-46980 València, Spain*

**Bingbing Liu** - *State Key Laboratory of Superhard Materials, International Centre of Future Science, Jilin University, Changchun 130012, China*

**Notes**

The authors declare no competing financial interest.

# ACKNOWLEDGMENTS

# REFERENCES

(1) Bousseksou, A.; Molnár, G.; Salmon, L.; Nicolazzi, W., Molecular spin crossover phenomenon: recent achievements and prospects. *Chem. Soc. Rev.* **2011,** *40* (6), 3313-3335.
(2) Munoz, M. C.; Real, J. A., Thermo-, piezo-, photo-and chemo-switchable spin crossover iron (II)-metallocyanate based coordination polymers. *Coord. Chem. Rev.* **2011,** *255* (17-18), 2068-2093.
(3) Gaspar, A. B.; Molnár, G.; Rotaru, A.; Shepherd, H. J., Pressure effect investigations on spin-crossover coordination compounds. *C. R. Chim.* **2018,** *21* (12), 1095-1120.
(4) Gakiya-Teruya, M.; Jiang, X.; Le, D.; Ungor, O.; Durrani, A. J.; Koptur-Palenchar, J. J.; Jiang, J.; Jiang, T.; Meisel, M. W.; Cheng, H.-P., Asymmetric Design of Spin-Crossover Complexes to Increase the Volatility for Surface Deposition. *J. Am. Chem. Soc.* **2021,** *143* (36), 1456314572.
(5) Gütlich, P.; Hauser, A.; Spiering, H., Thermal and optical switching of iron (II) complexes. *Angew. Chem., Int. Ed. Engl.* **1994,** *33* (20), 2024-2054.
(6) Levchenko, G.; Khristov, A.; Varyukhin, V., Spin crossover in iron (II)-containing complex compounds under a pressure. *Low Temp. Phys.* **2014,** *40* (7), 571-585.
(7) Coronado, E.; Giménez-López, M. C.; Levchenko, G.; Romero, F. M.; García-Baonza, V.; Milner, A.; Paz-Pasternak, M., Pressure-tuning of magnetism and linkage isomerism in iron (II) hexacyanochromate. *J. Am. Chem. Soc.* **2005,** *127* (13), 4580-4581.
(8) Levchenko, G.; Bukin, G.; Fylymonov, H.; Li, Q.; Gaspar, A. B. n.; Real, J. A., Electrical voltage control of the pressure-induced spin transition at room temperature in the microporous 3D polymer [Fe (pz) Pt (CN) $_4$]. *J. Phys. Chem. C* **2019,** *123* (9), 5642-5646.
(9) Martínez, V.; Gaspar, A. B.; Muñoz, M. C.; Bukin, G. V.; Levchenko, G.; Real, J. A., Synthesis and Characterisation of a New Series of Bistable Iron (II) Spin‐Crossover 2D Metal–Organic Frameworks. *Chem. - Eur. J.* **2009,** *15* (41), 10960-10971.
(10) Gaspar, A.; Agustí, G.; Martínez, V.; Muñoz, M.; Levchenko, G.; Real, J. A., Spin crossover behaviour in the iron (II)-2, 2-dipyridilamine system: Synthesis, X-ray structure and magnetic studies. *Inorg. Chim. Acta.* **2005,** *358* (13), 4089-4094.
(11) Doistau, B.; Benda, L.; Hasenknopf, B.; Marvaud, V.; Vives, G., Switching magnetic properties by a mechanical motion. *Magnetochemistry* **2018,** *4* (1), 5.
(12) Babilotte, K.; Boukheddaden, K., Theoretical investigations on the pressure effects in spin-crossover materials: Reentrant phase transitions and other behavior. *Phys. Rev. B.* **2020,** *101* (17), 174113.
(13) Zoppellaro, G.; Tucek, J.; Ugolotti, J.; Aparicio, C.; Malina, O.; Čépe, K. r.; Zbořil, R., Triggering two-step spin bistability and large hysteresis in spin crossover nanoparticles via molecular nanoengineering. *Chem. Mater.* **2017,** *29* (20), 8875-8883.
(14) Benaicha, B.; Van Do, K.; Yangui, A.; Pittala, N.; Lusson, A.; Sy, M.; Bouchez, G.; Fourati, H.; Gómez-García, C. J.; Triki, S., Interplay between spin-crossover and luminescence in a multifunctional single crystal iron (ii) complex: towards a new

(15) Santoro, A.; Kershaw Cook, L. J.; Kulmaczewski, R.; Barrett, S. A.; Cespedes, O.; Halcrow, M. A., Iron (II) complexes of tridentate indazolylpyridine ligands: Enhanced spin-crossover hysteresis and ligand-based fluorescence. *Inorg. Chem.* **2015,** *54* (2), 682-693.
(16) Ksenofontov, V.; Levchenko, G.; Spiering, H.; Gütlich, P.; Létard, J.-F.; Bouhedja, Y.; Kahn, O., Spin crossover behavior under pressure of Fe (PM-L)$_2$ (NCS)$_2$ compounds with substituted 2′-pyridylmethylene 4-anilino ligands. *Chem. Phys. Lett.* **1998,** *294* (6), 545-553.
(17) Gaspar, A. B.; Levchenko, G.; Terekhov, S.; Bukin, G.; Valverde-Muñoz, J.; Muñoz-Lara, F. J.; Seredyuk, M.; Real, J. A., The effect of pressure on the cooperative spin transition in the 2D coordination polymer {Fe (phpy)$_2$ [Ni (CN)$_4$]}. *Eur. J. Inorg. Chem.* **2014,** *2014*, 429-433.
(18) Levchenko, G.; Gaspar, A. B.; Bukin, G.; Berezhnaya, L.; Real, J. A., Pressure effect studies on the spin transition of microporous 3D polymer [Fe (pz) Pt (CN)$_4$]. *Inorg. Chem.* **2018,** *57* (14), 8458-8464.
(19) Galet, A.; Gaspar, A. B.; Muñoz, M. C.; Levchenko, G.; Real, J. A., Pressure Effect and Crystal Structure Reinvestigations on the Spin Crossover System: [Fe (bt)$_2$ (NCS)$_2$] (bt= 2, 2 '-Bithiazoline) Polymorphs A and B. *Inorg. Chem.* **2006,** *45* (24), 9670-9679.
(20) Levchenko, G.; Bukin, G.; Gaspar, A.; Real, J., The pressure-induced spin transition in the Fe (phen)$_2$ (NCS)$_2$ model compound. *Russ. J. Phys. Chem. A.* **2009,** *83* (6), 951-954.
(21) Yuan, M.; Levchenko, G.; Li, Q.; Berezhnaya, L.; Fylymonov, H.; Gaspar, A. B.; Seredyuk, M.; Real, J. A., Variable cooperative interactions in the pressure and thermally induced multistep spin transition in a two-dimensional iron (II) coordination polymer. *Inorg. Chem.* **2020,** *59* (15), 10548-10556.
(22) Decurtins, S.; Gütlich, P.; Köhler, C. P.; Spiering, H.; Hauser, A., Light-induced excited spin state trapping in a transition-metal complex: The hexa-1-propyltetrazole-iron (II) tetrafluoroborate spin-crossover system. *Chem. Phys. Lett.* **1984,** *105* (1), 1-4.
(23) Létard, J.-F., Photomagnetism of iron (II) spin crossover complexes—the T (LIESST) approach. *J. Mater. Chem.* **2006,** *16* (26), 2550-2559.
(24) Létard, J.-F.; Guionneau, P.; Rabardel, L.; Howard, J. A.; Goeta, A. E.; Chasseau, D.; Kahn, O., Structural, magnetic, and photomagnetic studies of a mononuclear iron (II) derivative exhibiting an exceptionally abrupt spin transition. Light-induced thermal hysteresis phenomenon. *Inorg. Chem.* **1998,** *37* (17), 4432-4441.
(25) Hauser, A., Light-induced spin crossover and the high-spin→ low-spin relaxation. *Spin crossover in transition metal compounds II* **2004**, 155-198.
(26) Brady, C.; McGarvey, J. J.; McCusker, J. K.; Toftlund, H.; Hendrickson, D. N., Time-resolved relaxation studies of spin crossover systems in solution. *Spin Crossover in Transition Metal Compounds III* **2004**, 1-22.
(27) Bonhommeau, S.; Molnár, G.; Galet, A.; Zwick, A.; Real, J. A.; McGarvey, J. J.; Bousseksou, A., One shot laser pulse induced reversible spin transition in the spin‐crossover complex [Fe (C$_4$H$_4$N$_2$){Pt (CN)$_4$}] at room temperature. *Angew. Chem. Int. Ed.* **2005,** *117* (26), 4137-4141.
(28) Dong, X.; Lorenc, M.; Tretyakov, E. V.; Ovcharenko, V. I.; Fedin, M. V., Light-induced spin state switching in copper (ii)-nitroxide-based molecular magnet at room temperature. *J. Phys. Chem. Lett.* **2017,** *8* (22), 5587-5592.
(29) Rohlf, S.; Gruber, M.; Flöser, B. M.; Grunwald, J.; Jarausch, S.; Diekmann, F.; Kalläne, M.; Jasper-Toennies, T.; Buchholz, A.; Plass, W., Light-induced spin crossover in an Fe (II) low-spin complex enabled by surface adsorption. *J. Phys. Chem. Lett.* **2018,** *9* (7), 1491-1496.
(30) Goujon, A.; Gillon, B.; Gukasov, A.; Jeftic, J.; Nau, Q.; Codjovi, E.; Varret, F., Photoinduced molecular switching studied by polarized neutron diffraction. *Phys. Rev. B.* **2003,** *67* (22), 220401.
(31) Bousseksou, A.; Varret, F.; Goiran, M.; Boukheddaden, K.; Tuchagues, J.-P., The spin crossover phenomenon under high magnetic field. *Spin Crossover in Transition Metal Compounds III* **2004**, 65-84.
(32) Davesne, V.; Gruber, M.; Miyamachi, T.; Da Costa, V.; Boukari, S.; Scheurer, F.; Joly, L.; Ohresser, P.; Otero, E.; Choueikani, F., First glimpse of the soft x-ray induced excited spin-state trapping effect dynamics on spin cross-over molecules. *J. Chem. Phys.* **2013,** *139* (7), 074708.
(33) Gaspar, A. B.; Muñoz, M. C.; Moliner, N.; Ksenofontov, V.; Levchenko, G.; Gütlich, P.; Real, J. A., Polymorphism and pressure driven thermal spin crossover phenomenon in [Fe (abpt)$_2$ (NC X)$_2$](X= S, and Se): Synthesis, structure and magnetic properties. *Monatsh. Chem.* **2003,** *134* (2), 285-294.
(34) Ohba, M.; Yoneda, K.; Agustí, G.; Munoz, M. C.; Gaspar, A. B.; Real, J. A.; Yamasaki, M.; Ando, H.; Nakao, Y.; Sakaki, S., Bidirectional chemo‐switching of spin state in a microporous framework. *Angew. Chem. Int. Ed.* **2009,** *48* (26), 4767-4771.
(35) Southon, P. D.; Liu, L.; Fellows, E. A.; Price, D. J.; Halder, G. J.; Chapman, K. W.; Moubaraki, B.; Murray, K. S.; Létard, J.-F.; Kepert, C. J., Dynamic interplay between spin-crossover and host– guest function in a nanoporous metal– organic framework material. *J. Am. Chem. Soc.* **2009,** *131* (31), 10998-11009.
(36) Valverde-Muñoz, F. J.; Seredyuk, M.; Muñoz, M. C.; Znovjyak, K.; Fritsky, I. O.; Real, J. A., Strong Cooperative Spin Crossover in 2D and 3D FeII-MI, II Hofmann-Like Coordination Polymers Based on 2-Fluoropyrazine. *Inorg. Chem.* **2016,** *55* (20), 10654-10665.
(37) Li, R.; Levchenko, G.; Valverde-Muñoz, F. J.; Gaspar, A. B.; Ivashko, V. V.; Li, Q.; Liu, B.; Yuan, M.; Fylymonov, H.; Real, J. A., Pressure Tunable Electronic Bistability in Fe (II) Hofmann-like Two-Dimensional Coordination Polymer [Fe (Fpz)$_2$Pt (CN)$_4$]: A Comprehensive Experimental and Theoretical Study. *Inorg. Chem.* **2021,** *60* (21), 16016-16028.

(38) Molnár, G.; Niel, V.; Real, J.-A.; Dubrovinsky, L.; Bousseksou, A.; McGarvey, J. J., Raman Spectroscopic Study of Pressure Effects on the Spin-Crossover Coordination Polymers Fe (Pyrazine)[M (CN)$_4$]⊙ 2H2O (M= Ni, Pd, Pt). First Observation of a Piezo-Hysteresis Loop at Room Temperature. *J. Phys. Chem. B.* **2003,** *107* (14), 3149-3155.

(39) Lada, Z. G.; Andrikopoulos, K. S.; Polyzou, C. D.; Tangoulis, V.; Voyiatzis, G. A., Monitoring the spin crossover phenomenon of [Fe (2‐mpz)$_2$Ni (CN)$_4$] 2D Hofmann‐type polymer nanoparticles via temperature‐dependent Raman spectroscopy. *J. Raman Spectrosc.* **2020,** *51* (11), 2171-2181.

(40) Levchenko, G.; Bukin, G.; Terekhov, S.; Gaspar, A.; Martinez, V.; Munoz, M.; Real, J., Pressure-induced cooperative spin transition in iron (II) 2D coordination polymers: Room-temperature visible spectroscopic study. *J. Phys. Chem. B.* **2011,** *115* (25), 8176-8182.

(41) Galet, A.; Gaspar, A. B.; Muñoz, M. C.; Bukin, G. V.; Levchenko, G.; Real, J. A., Tunable Bistability in a Three‐Dimensional Spin‐Crossover Sensory‐and Memory‐Functional Material. *Adv. Mater.* **2005,** *17* (24), 2949-2953.

(42) Kalita, V.; Levchenko, G., The average value of the spin squared operator as an order parameter for spin phase transitions without spontaneous lowering of symmetry. *J. Phys. Commun.* **2020,** *4* (9), 095024.

(43) Slichter, C.; Drickamer, H., Pressure‐induced electronic changes in compounds of iron. *J. Chem. Phys.* **1972,** *56* (5), 2142-2160.

(44) Chernyshov, D.; Bürgi, H.-B.; Hostettler, M.; Törnroos, K. W., Landau theory for spin transition and ordering phenomena in Fe (II) compounds. *Phys. Rev. B.* **2004,** *70* (9), 094116.

(45) Levchenko, G.; Khristov, A.; Kuznetsova, V.; Shelest, V., Pressure and temperature induced high spin–low spin phase transition: Macroscopic and microscopic consideration. *J. Phys. Chem. Solids.* **2014,** *75* (8), 966-971.

(46) Drickamer, H.; Frank, C.; Slichter, C., Optical versus thermal transitions in solids at high pressure. *Proc. Natl. Acad. Sci. U.S.A.* **1972**, 69 (4), 933-937.

(47) Rusu, I.; Manolache-Rusu, I. C.; Diaconu, A.; Palamarciuc, O.; Gural'skiy, I. y. A.; Molnar, G.; Rotaru, A., Pressure gradient effect on spin-crossover materials: Experiment vs theory. *J. Appl. Phys.* **2021**, 129 (6), 06450